\documentclass[reprint,a4paper,aps,prb]{revtex4-2}
\pdfoutput=1

\usepackage{graphicx}
\usepackage{amsmath}
\usepackage{amsfonts}
\usepackage{physics}
\usepackage{bm}
\usepackage{xcolor}
\usepackage{hyperref}
\usepackage{textpos}

\newcommand{\mrm}{\mathrm}

\newcommand{\Psit}{\Psi_{\bm{\theta}}}
\newcommand{\bsigma}{\bm{\sigma}}
\newcommand{\btheta}{\bm{\theta}}

\usepackage{tikz}
\newcommand{\cbr}[1]{%
  \raisebox{0.2ex}{%
    \begin{tikzpicture}[scale=0.12]
      \node[anchor=east] at (-0.2, -1.5) {\scriptsize $r +$};
      \foreach \x in {0,1,2} {
        \foreach \y in {0,1,2} {
          \fill[gray!20] (\x,-\y) rectangle ++(1,-1);
        }
      }
      \ifnum#1=00 \fill[black] (0,0) rectangle ++(1,-1);\fi
      \ifnum#1=01 \fill[black] (0,-1) rectangle ++(1,-1);\fi
      \ifnum#1=02 \fill[black] (0,-2) rectangle ++(1,-1);\fi
      \ifnum#1=10 \fill[black] (1,0) rectangle ++(1,-1);\fi
      \ifnum#1=11 \fill[black] (1,-1) rectangle ++(1,-1);\fi
      \ifnum#1=12 \fill[black] (1,-2) rectangle ++(1,-1);\fi
      \ifnum#1=20 \fill[black] (2,0) rectangle ++(1,-1);\fi
      \ifnum#1=21 \fill[black] (2,-1) rectangle ++(1,-1);\fi
      \ifnum#1=22 \fill[black] (2,-2) rectangle ++(1,-1);\fi
      \draw (0,0) grid (3,-3);
    \end{tikzpicture}%
  }%
}

\newcommand{\cb}[1]{%
  \raisebox{0.2ex}{%
    \begin{tikzpicture}[scale=0.12]
      \foreach \x in {0,1,2} {
        \foreach \y in {0,1,2} {
          \fill[gray!20] (\x,-\y) rectangle ++(1,-1);
        }
      }
      \ifnum#1=00 \fill[black] (0,0) rectangle ++(1,-1);\fi
      \ifnum#1=01 \fill[black] (0,-1) rectangle ++(1,-1);\fi
      \ifnum#1=02 \fill[black] (0,-2) rectangle ++(1,-1);\fi
      \ifnum#1=10 \fill[black] (1,0) rectangle ++(1,-1);\fi
      \ifnum#1=11 \fill[black] (1,-1) rectangle ++(1,-1);\fi
      \ifnum#1=12 \fill[black] (1,-2) rectangle ++(1,-1);\fi
      \ifnum#1=20 \fill[black] (2,0) rectangle ++(1,-1);\fi
      \ifnum#1=21 \fill[black] (2,-1) rectangle ++(1,-1);\fi
      \ifnum#1=22 \fill[black] (2,-2) rectangle ++(1,-1);\fi
      \draw (0,0) grid (3,-3);
    \end{tikzpicture}%
  }%
}

\begin{document}
\title{Design principles of deep translationally-symmetric neural quantum states for frustrated magnets}

\author{Rajah P. Nutakki}
\affiliation{CPHT, CNRS, École Polytechnique, Institut Polytechnique de Paris, 91120 Palaiseau, France.}
\affiliation{Coll\`ege de France, Universit\'e PSL, 11 place Marcelin Berthelot, 75005 Paris, France}
\affiliation{Inria Paris-Saclay, 91120 Palaiseau, France}
\affiliation{LIX, CNRS, Ecole Polytechnique, Institut Polytechnique de Paris, 91120 Palaiseau, France}
\author{Ahmedeo Shokry}
\affiliation{CPHT, CNRS, École Polytechnique, Institut Polytechnique de Paris, 91120 Palaiseau, France.}
\affiliation{Coll\`ege de France, Universit\'e PSL, 11 place Marcelin Berthelot, 75005 Paris, France}
\affiliation{Inria Paris-Saclay, 91120 Palaiseau, France}
\affiliation{LIX, CNRS, Ecole Polytechnique, Institut Polytechnique de Paris, 91120 Palaiseau, France}
\author{Filippo Vicentini}
\affiliation{CPHT, CNRS, École Polytechnique, Institut Polytechnique de Paris, 91120 Palaiseau, France.}
\affiliation{Coll\`ege de France, Universit\'e PSL, 11 place Marcelin Berthelot, 75005 Paris, France}
\affiliation{Inria Paris-Saclay, 91120 Palaiseau, France}
\affiliation{LIX, CNRS, Ecole Polytechnique, Institut Polytechnique de Paris, 91120 Palaiseau, France}

%\date{xx.xx.2025}

\begin{abstract}
Deep neural network quantum states have emerged as a leading method for studying the ground states of quantum magnets.
Successful architectures exploit translational symmetry, but the root of their effectiveness and differences between architectures remain unclear.
Here, we apply the ConvNext architecture, designed to incorporate elements of transformers into convolutional networks, to quantum many-body ground states.
We find that it is remarkably similar to the factored vision transformer, which has been employed successfully for several frustrated spin systems, allowing us to relate this architecture to more conventional convolutional networks.
Through a series of numerical experiments we design the ConvNext to achieve greatest performance at lowest computational cost, then apply this network to the Shastry-Sutherland and $J_1-J_2$ models, obtaining variational energies comparable to the state of the art, providing a blueprint for network design choices of translationally-symmetric architectures to tackle challenging ground-state problems in frustrated magnetism. 
\end{abstract}

% insert suggested keywords - APS authors don't need to do this
%\keywords{}

%\maketitle must follow title, authors, abstract, and keywords
\maketitle
\section{Introduction}
Investigating the ground state properties of frustrated quantum magnets remains a formidable challenge in computational physics, yet crucial for understanding what states of matter can be experimentally realized in solid-state and synthetic quantum platforms.
Several methods have been developed over the years, but have their limitations.
Tensor networks ~\cite{white1992,white1993,schollwock2011,stoudenmire2012,ran2020}, for example, are a reliable tool when an efficient contraction scheme exists, while Quantum Monte Carlo techniques~\cite{prokofev1998d,sandvik1999,syljuasen2002,evertz2003,houcke2008,becca2017} are generally well suited for investigating the finite temperature properties of sign-problem-free Hamiltonians.

Neural quantum states (NQS)~\cite{carleo2017a} have recently emerged as a promising family of variational wave-functions for use within Variational Monte Carlo (VMC)~\cite{becca2017}, achieving state-of-the-art accuracy for 2D frustrated quantum spin systems~\cite{viteritti2024,duric2024}, such as the $J_1-J_2$ model on the square lattice~\cite{chen2024a,rende2023,roth2023a,li2022,liang2023} and offering promising results for the calculation of their dynamical properties~\cite{gravina2024,Mauron2024DynamicsTopology,MendesSantos2023Spectral}.
Their recent success stems from the ability to optimize large networks with many variational parameters using Stochastic Reconfiguration~\cite{chen2024a,sorella1998,Amari1998NGD,becca2017,Martens2020NGD,Dash2025QGT}, which is closely connected to imaginary time evolution.

Many neural-network wave-functions rely on the same or slightly adapted \textit{encoder} architectures used in several domains of Machine Learning, with only minor modifications to the final layers in order to output complex-valued amplitudes instead of a set of classification probabilities or text.
Two succesful encoder architectures, used to obtain the most accurate results for variational ground-state calculations, are the convolutional residual network (ResNet)~\cite{chen2024a} and the vision transformer with factored attention (fViT)~\cite{rende2024b,viteritti2023,rende2023,viteritti2024}.
Both are built out of translationally-equivariant blocks, allowing one to efficiently enforce translational symmetries in the variational wavefunction, which appears to be instrumental in the success of the method. 
Evidence is emerging that more general attention mechanisms which are not translationally-equivariant do not (significantly) improve performance~\cite{rende2024a, chen2025}.
However, the similarities and differences between the fViT and ResNet architectures have not been investigated, and it remains unclear how best to exploit their properties when studying different frustrated spin systems.

In this work, we implement the ConvNext~\cite{liu2022,woo2023} as an NQS, an architecture designed to elevate the performance of convolutional neural networks (CNNs) to that of transformers for computer vision tasks.
Our motivation is to harness the power of transformers~\cite{vaswani2017,islam2024}, whilst retaining the ability to efficiently impose translational symmetries, as well as the locality inherent in convolutional networks.
Remarkably, we find that the ConvNext is similar to the fViT ~\cite{rende2024b} (and can be made almost equivalent by specific choices of hyperparameters), showing the convergence of two directions in network design: making a CNN more transformer-like, or making the attention in a ViT translationally-equivariant.
Thus the ConvNext offers a way to understand translationally-symmetric architectures in a clear and unified manner.
First, we summarize the properties of translationally-symmetric NQS (section~\ref{sec:tenqs}), putting their components into the embedding-encoder-readout framework of Ref.~\cite{viteritti2024}.
We then perform a series of numerical experiments (section~\ref{sec:experiments}) on the ConvNext clarifying the effect of patching, network depth, kernel width and the role of the readout block in light of this framework.
Using our optimal design, we obtain variational energies for two benchmark problems, the $J_1-J_2$ model and Shastry-Sutherland which are comparable to the state of the art with NQS.
We end by discussing (section~\ref{sec:discussion}) outstanding questions on translationally-symmetric NQS and comparing to other types of architectures.
Our work makes clear the design choices when employing translationally-symmetric NQS for ground states of frustrated spin systems, such that they can be employed to tackle challenging problems in frustrated magnetism and beyond.

\section{The Shastry-Sutherland Model}
\label{sec:ham}
\begin{figure}
\centering
    \includegraphics[width=5cm]{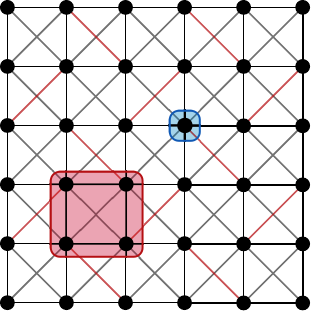}
    \caption{The $J_1$ (black) and $J_2$ couplings of the (red) Shastry-Sutherland and (grey and red) $J_1-J_2$ models.
    The sites in the unit cell are highlighted for the Shastry-Sutherland (red) and $J_1-J_2$ (blue) models.
    }
    \label{fig:shastry}
\end{figure}
\begin{figure*}[ht]
	\centering
	\includegraphics[width=\textwidth]{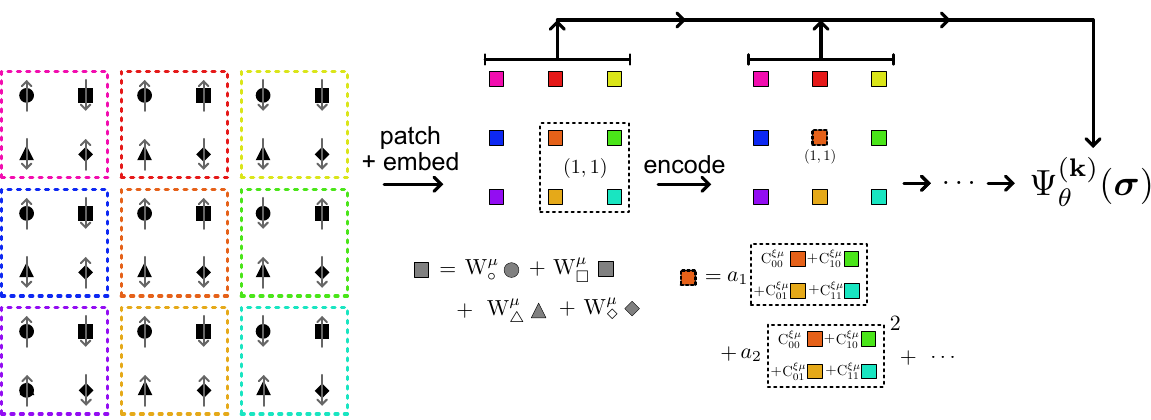}
	\caption{Illustration of a patched translationally-symmetric NQS. 
	The patching and embedding of the input means the remainder of the network operates on sublattice-dependent quantities, a useful bias for learning the sign structure of ground states.
	Convolution-like encoding blocks then construct a hidden representation out of (high-order) correlators of these quantities, before projection to a momentum, $\mathbf{k}$, and readout of a complex amplitude.}
	\label{fig:general_structure}
\end{figure*}

We use the Shastry-Sutherland model~\cite{sriramshastry1981} as our benchmark to evaluate performance of different NQS architecture choices.
We make this choice as the model has a multi-site unit cell, like many in frustrated magnetism~\cite{lacroix2011}, and was recently studied using the fViT, so can compare with Ref.~\cite{viteritti2024}.
We also evaluate the performance of our final network design on the related $J_1-J_2$ model, allowing us to compare to many other approaches which have been applied to the same model.

Both models are defined by the $S = 1/2$ Hamiltonian
\begin{equation}
    \hat{H} = J_1 \sum_{\langle ij \rangle} \hat{\mathbf{S}}_i \cdot \hat{\mathbf{S}}_j +J_2 \sum_{(ij) \in J_2} \hat{\mathbf{S}}_i \cdot \hat{\mathbf{S}}_j
    \label{eq:shastry}
\end{equation}
where $\hat{\mathbf{S}}_i = (\hat{S}^x_i, \hat{S}^y_i, \hat{S}^z_i) = 1/2(\hat{\sigma}^x_i, \hat{\sigma}^y_i, \hat{\sigma}^z_i)$ with variables on a square grid.
The $J_1$ term couples nearest neighbour sites, whereas $J_2$ couples all ($J_1-J_2$ model) or 1/4 (Shastry-Sutherland) of next nearest neighbours, as shown in Fig.~\ref{fig:shastry}.

Both the Shastry-Sutherland~\cite{yang2022,wang2022a,keles2022,viteritti2024} and $J_1-J_2$\cite{wang2018a,ferrari2020a,nomura2021a,liu2022a} models are thought to host a $\mathbb{Z}_2$ Dirac spin liquid~\cite{hu2013,ferrari2018b} ground state, which were recently proposed to be part of the same phase~\cite{maity2024}.
The ground state in these parameter regimes is long-range correlated, has a non-trivial sign structure, as well as nearby competing phases, making computations challenging and therefore a useful benchmark.

For the remainder of this manuscript we will restrict ourselves to $J_1=1$, $J_2 = 0.8$, of the Shastry-Sutherland and $J_1 = 1$, $J_2 = 0.5$ of the $J_1-J_2$ model, within the putative spin liquid phase, taking the model on an $L \times L$ square geometry with periodic boundary conditions.

\section{Translationally-Symmetric Neural Quantum States}
\label{sec:tenqs}

\paragraph{Sign structure and embedding:}
A major hurdle for NQS-based VMC applications to frustrated (or fermionic) ground states is learning their sign structure~\cite{choo2019,szabo2020}, which can even lead to an apparent exponential complexity emerging if a proper prior is not used~\cite{Denis2025Comment}.
While a clear cause is unknown, it might be related to a vanishing amplitudes of the wave-function, which increase the variance of the gradient estimator, emerging when the optimization attempts to flip the sign~\cite{Sinibaldi2023}.
For models on bipartite lattices, the sign structure (Marshall sign rule) is known exactly~\cite{marshall1955}.
For non-bipartite systems, like the $J_1$–$J_2$ model, the sign rule may still serve as a good approximate prior~\cite{szabo2020,roth2023a}, however for a general lattice this structure is not analytically known.

Starting from a wavefunction that already captures a reasonable approximation to the sign structure, either through prior knowledge or inductive bias, can significantly improve optimization and lead to more accurate ground-state calculations. 
For example, when signs depend on sublattice-dependent quantities, as is the case in models which (approximately) obey the Marshall sign rule, it is advantageous to design the network to operate on these quantities.

This can be done using an \textit{embedding} layer at the beginning of the network, which partitions the lattice into $P$ patches, with each of the sites within a patch assigned to a separate sublattice, as illustrated in Fig.\ref{fig:general_structure}.
We denote the input configuration in the computational basis as $\sigma_{\mathbf{r}, x}$, where $\mathbf{r}$ labels the position of the patch (the dashed squares in Fig.\ref{fig:general_structure}) and $x$ labels the sites (sublattices) within the patch. 
The embedding layer $f^s$ then applies a sublattice-dependent linear transformation $\mrm W^\mu_{x}$ to the spin variables within each patch
This embedding is written as (suppressing vector notation for clarity):
\begin{equation}
	\tilde{\sigma}_{r}^{\mu} = f^s(\sigma)_{r}^{\mu} =  \sum_x \mrm W^{\mu}_x \sigma_{r,x},
	\label{eq:stem_hidden}
\end{equation}
where $\mu$ indexes the embedding dimension (features), of total size $N_f$. 
Because this transformation explicitly acts on the internal sublattice structure of each patch, it naturally biases the network to learn the sign structure of the ground state on a bipartite lattice, discussed further in Appendix~\ref{app:sublattice}.

In networks which do not include this embedding, it has been necessary to provide the sign rule as a prior to facilitate training and avoid convergence issues~\cite{chen2024a}.

\paragraph{Symmetries and encoder:}
The core of a translationally-symmetric NQS is the encoder, which processes the variables $\tilde\sigma_r^\mu$ through a series of translationally-equivariant transformations,
\begin{equation}
    \tilde{\bm{\sigma}}_{\hat{t} r}[l+1] = g^e(t^{-1} \tilde{\bm{\sigma}}_r[l]),
    \label{eq:trans_equiv}
\end{equation}
where $g^e$ is the encoder function, $t$ is an element of the translation group, $T$, $l$ labels the layers of the encoder and $\tilde{\bm{\sigma}}_r =  (\tilde{\sigma}_r^{0}, \dots,\tilde{\sigma}_r^{N_f-1})$ ~\cite{roth2023a}\footnote{The translation of the input gives a corresponding translation of the output}.
This encoder is typically built from convolutional layers, combined with nonlinearities and skip connections.

The network can be later made translationally-symmetric (up to the level of the patches) by projecting to a specific momentum, $\mathbf{k}$,
\begin{equation}
    \Psi_{\theta}^{(\mathbf{k})}(\bm{\sigma}) = \frac{1}{\abs{T}} \sum_{t \in T} e^{-i \mathbf{k} \cdot \mathbf{r}_t} \Psi_{\theta}(t^{-1} \bm{\sigma}),
    \label{eq:trans_symm}
\end{equation}
from a single evaluation of the network, where $\mathbf{r}_t$ is the shift in coordinates corresponding to $t$~\footnote{The encoder is not equivariant, and the whole network not translationally-symmetric, under sub-patch translations. For example, considering the square lattice with $2\times 2$ patches of Fig.~\ref{fig:patching}, the network would not transform correctly under translations by 1 site along the $x$ or $y$ axes}. 
Point group and spin symmetries can be included with an overhead of $\abs{G}$ network evaluations (Eq.~\ref{eq:symmetrize}), where $\abs{G}$ is the number of elements in the corresponding group.

From a physical point of view, the encoder can be thought of as extracting $\sigma_i^z$ correlators of all orders (see Appendix~\ref{app:correlators}).
At each layer, the patched features $\tilde\sigma_r^\mu[l]$ are updated according to:
\begin{multline}
	 \tilde{\bm{\sigma}}_{r}[l+1] =   \tilde{\bm \sigma}_{r}[l] + \\  g^e \bigg( \tilde{\bm{\sigma}}_{\cbr{00}}[l],
     \tilde{\bm{\sigma}}_{\cbr{10}}[l],
     \dots, 
     \tilde{\bm{\sigma}}_{\cbr{22}}[l]\bigg),
	 \label{eq:encoder_recursion}
\end{multline}
% where $r+\{\cb{00}, \cb{01}, \dots \}$ are the neighboring patches within the receptive field $\cb{33}$ \ of patch $r$.
where $\tilde{\bm \sigma}_{r}[l]$ denotes the feature vector at patch $r$ in layer $l$ of the encoder, corresponding to the residual connection. The updated feature at the next layer, {$\tilde{\bm{\sigma}}_{r}[l+1]$,} is computed as the sum of the current feature and a contribution from the neighboring patches within the receptive field. These neighbors are located at relative offsets  $\{\cb{00}, \cb{01}, \dots \}$ from patch $r$. In essence, the new representation at patch $r$ aggregates information from its local neighborhood while retaining its original feature.
The skip connection ensures that the original features (sublattice-dependent magnetizations) are preserved and passed along, enabling the network to encode higher-order structure on top without foregoing lower-order contributions.
The recursive structure allows the encoder to build up combinations of the input variables, including products of $\sigma_i^z$ values at different sites, generating multi-spin correlators across the lattice.

The depth of the encoder and the size of the receptive field determine how nonlocal and how many different spin variables these correlators can contain.
The encoder uses these correlators to construct a hidden representation, which a final readout layer aggregates, projecting to a desired momentum sector (Eq.~\ref{eq:trans_symm}) and producing a complex scalar amplitude. 

These elements can be broadly organized into an embedding-encoder-readout structure~\cite{viteritti2024} (see e.g Fig.~\ref{fig:convnext}), which we find useful for discussing the various properties of the network in this article. 

\subsection{ConvNext Quantum State}
\begin{figure}[t]
\centering
\includegraphics[width=6cm]{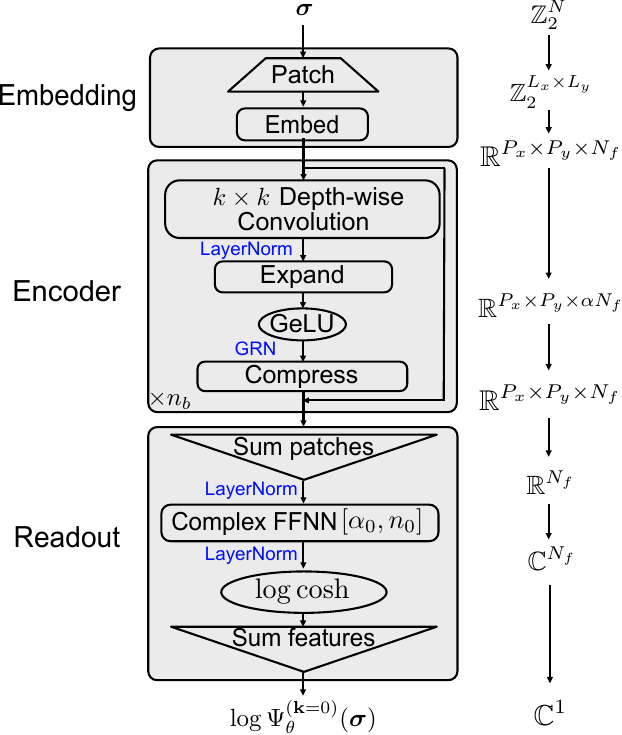}
    \caption{The ConvNext architecture adapted for use as an NQS. The right-hand side displays the shape and type of the data as it passes through the network.
    The number of features, $N_f$, expansion ratio, $\alpha$, number of encoder blocks, $n_b$, as well as convolutional kernel width, $k$ and size ($n_o, \alpha_o$) of the feedforward neural network (FFNN) in the readout block are all hyperparameters.
    The number of patches along the Cartesian axes are denoted $P_x, P_y$.
    The embedding and encoder have real-valued parameters, with a complex-valued FFNN in the readout head resulting in a complex output.
    }
    \label{fig:convnext}
\end{figure}
In this section, we discuss an adaptation of the ConvNext architecture, originally introduced for computer vision tasks~\cite{liu2022,woo2023}, to represent quantum many-body states (see Fig.~\ref{fig:convnext}). 
The ConvNext combines structural elements of CNNs and transformers, and its use as an NQS provides valuable insights into designing translationally-symmetric neural architectures.

The original ConvNext architecture~\cite{liu2022} consists of multiple encoding stages, each progressively patching the input into smaller spatial dimensions while increasing the number of feature channels.
As a result, the network outputs a highly compact representation with rich feature content at each spatial location. 
However, each additional patching stage partially breaks translational symmetry within the resulting patches, a trade-off typically acceptable or even beneficial in general machine learning applications~\cite{wilson2025}.

For neural quantum states, however, preserving exact lattice translational symmetry is essential for accurate calculations. 
We therefore restrict ourselves to a single-stage architecture, patching only one time and therefore losing only a small number of symmetries (if the lattice has a single-site unit cell) which could be restored later. 
Under this constraint, the ConvNext encoder closely resembles the fViT architecture. 
Nevertheless, the hierarchical multi-stage structure could become advantageous when dealing with very large quantum systems consisting of thousands of lattice sites\cite{sprague2024a}.
%%%%%%%%%%%%%
\begin{figure}[t]
	\centering
	\includegraphics[width=0.8\columnwidth]{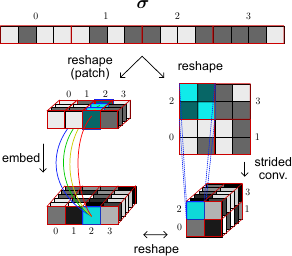}
	\caption{Two different ways of patching and embedding the input. The numbers label the patches, which are made up of the variables in red boxes. (Left) The data is first reshaped (patched) to $\mathbb{Z}_2^{P \times N_x}$, where $N_x$ is the number of sublattices created by the patching, before embedding to $\mathbb{R}^{P \times N_f}$. (Right) The data is reshaped to its lattice positions, before a strided convolution simultaneously performs the patching and embedding operations to $\mathbb{R}^{P_x \times P_y \times N_f}$. The two are mathematically equivalent.}
	\label{fig:stem}
\end{figure}
%%%%%%%%%%%%%

\paragraph{Patching and embedding:}

The first layer of a ConvNext stage is a stride-$s$ convolution which reduces the spatial dimension of the input by a factor of $s^2$ and increases the number of features~\cite{liu2022}. 
This operation is mathematically equivalent to first reshaping the input into  $s\times s$ patches and then applying a learnable embedding, as done in the embedding layer \ref{eq:stem_hidden} used by the fViT (see the illustration of Fig.~\ref{fig:stem}).
Here, we use the patch-and-embed implementation, which we find to be more efficient for the relatively small input sizes encountered in NQS.

\paragraph{Encoder:}
The ConvNext encoder transforms the embedded variables $\tilde{\sigma}_r^\mu$ through a series of convolutional blocks designed to efficiently encode spatial correlations.
Previous convolutional wavefunctions built on top of ResNets~\cite{chen2024a,schmitt2020} relied on layers of standard  convolutions that mix all spatial and feature degrees of freedom.
Those can be written as
\begin{equation}
	\tilde{\sigma}^{\xi}_{r}[l+1] = g\bigg(\sum_{\mu} \sum_{\delta} \mrm C^{\xi,\mu}_{\delta} \tilde{\sigma}_{r+\delta}^{\mu}[l] \bigg) + \tilde{\sigma}^{\xi}_{r}[l],
	\label{eq:convolution}
\end{equation}
where $\mrm C$ is the convolutional kernel (a $k \times k \times N_f^2$ filter for 2D lattices), and $g$ is some element-wise nonlinearity.
While effective, these convolutions are relatively costly, as their memory usage and computational requirements scale quadratically with both $k$ and $N_f$. 

The key insight behind the ConvNext~\cite{liu2022} is to reduce this computational burden by splitting the convolution into two simpler steps. 
First, a depth-wise convolution acts independently on each feature channel, mixing only spatial information:
\begin{equation}
\textrm{Conv}(\tilde{\sigma})^\xi_{r} = \sum_{\delta} \mrm K^\xi_\delta ,\tilde{\sigma}_{r+\delta}^\xi[l],
\label{eq:convolution-dw}
\end{equation}
where $\mrm K$ is a depth-wise kernel of size $k \times k \times N_f$.
This is followed by a feature-mixing nonlinear transformation $g(\cdot)$ which does not mix the spatial information. 
This nonlinearity is, in practice, implemented as a 2-layer multi-layer perceptron with a single nonlinearity,
\begin{equation}
g^\xi(\mathbf{\tilde {\bm{\sigma}}_r}) = \sum_{\nu=1}^{\alpha N_f} V^{\xi,\nu} \mathrm{GeLU}\left(\sum_{\mu=1}^{N_f}W^{\nu,\mu}\tilde{\sigma}^\mu_r\right),
\label{eq:mlp}
\end{equation}
with matrices $W$ and $V$ having dimensions $(\alpha N_f, N_f)$ and $(N_f, \alpha N_f)$, respectively, where $\alpha$ controls the hidden layer density (also known as the \textit{expansion factor}).
Thus, the overall encoder layer becomes:
\begin{equation}
\tilde{\sigma}^\xi_{r}[l+1] = \tilde\sigma^\xi_{r}[l] + g^\xi(\textrm{Conv}(\tilde{\bm \sigma}[l])_r).
\label{eq:convnext-layer}
\end{equation}
This factorization reduces the computational cost from quadratic to linear scaling in the embedding dimension, and it has been shown empirically that this does not substantially reduce the expressivity.
In practice, $\alpha$ is set to between $1$ and $4$ and $N_f$ is on the order of $10$s to $100$s.
Therefore, the vast majority of the variational parameters are found within this nonlinearity and not in the spatial convolutions.

\paragraph{Relationship to the fViT}
We remark that the ConvNext, being inspired by a vision transformer, shares with it the same nonlinear activation $g$.
The greatest difference between a ViT and a ConvNext is in the attention mechanism: a standard ViT would not use a linear convolution to mix different spatial (patch) degrees of freedom but instead a dense matrix that depends on the input.
Rende et al.~\cite{rende2024b} pointed out that this attention would not be translationally equivariant, and proposed to replace it with a \textit{translationally-equivariant factored attention} that restores this equivariance~\cite{rende2024b}, later showing further empirical evidence motivating this choice~\cite{rende2024c}.
As detailed in Appendix~\ref{app:fmha_comp}, this is mathematically equivalent to a depth-wise convolution (as in the ConvNext) when employing as many heads as features and choosing a kernel size spanning the entire system.

\paragraph{Readout}
Given the similarities between the ConvNext and fViT, we use a similar readout block to Ref.~\cite{viteritti2023}, first summing over positions (projecting to the $\mathbf{k} = 0$ sector) before applying a complex feed-forward network of depth $n_o$ and hidden layers of size $\alpha_o N_f$.
Note that this design does not allow one to project to $\mathbf{k} \neq 0$ sectors, so we also verify the performance of a modified version which does.

\section{Numerical Experiments}
\label{sec:experiments}
We perform a series of numerical experiments to identify the optimal design and hyperparameter choices for the ConvNext.
As a benchmark, we decide to use the ground state search of the Shastry-Sutherland model.
We use, where possible, the same optimization protocol and sampling procedure, as described in appendix~\ref{app:optimization}.

The main hyperparameters for the network are the depth of the encoder, $n_b$, kernel width of the convolution, $k$, and encoder expansion ratio, $\alpha$, as well as the depths, $n_o$, and expansion factor, $\alpha_o$, in the readout.
In order to make our results comparable to those in~\cite{viteritti2024}, we fix $N_f = 72$.
We specify a choice of hyperparameters with the notation $(n_b,k,\alpha)[n_o,\alpha_o]$.
Since the unit cell of the model is a $2 \times 2$ plaquette (fig.~\ref{fig:shastry}), we use these as the patches, so that we can easily enforce the translational symmetry of the model.
\subsection{Patching}
%%%%%%
\begin{figure}
    \centering
    \hspace{-1cm}
    \includegraphics[width=8cm]{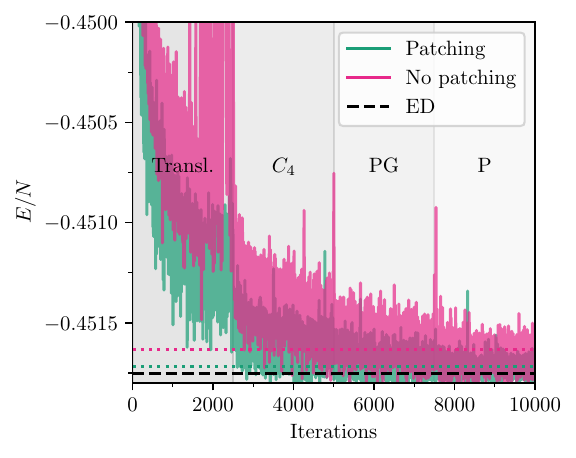}
    \caption{Optimization of the $(4,3, 2)[1,1]$ ConvNext with and without patching for $L = 6$.
    The respective final energies ($E/N = -0.451717(6)$ and $-0.451633(6)$) are shown with dotted lines and the ground state energy $E_{\mrm{GS}}/N = -0.4517531$~\cite{viteritti2024} dashed. 
    Relative errors along with measures of the computational cost are shown in fig.~\ref{fig:summary}.
    The annotations refer to the symmetrization stages of the optimization (see Appendix~\ref{app:optimization}).
    }
    \label{fig:patching}
\end{figure}

%%%%%%
%%%%%
\begin{figure*}
    \centering
    \begin{minipage}{6.5cm}
    \includegraphics[width=\textwidth]{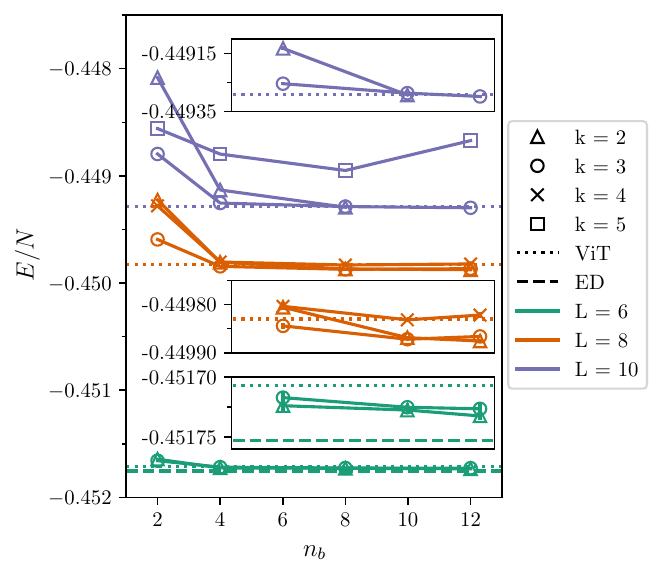}
    \end{minipage}
    \begin{minipage}{5cm}
    \includegraphics[width=\textwidth]{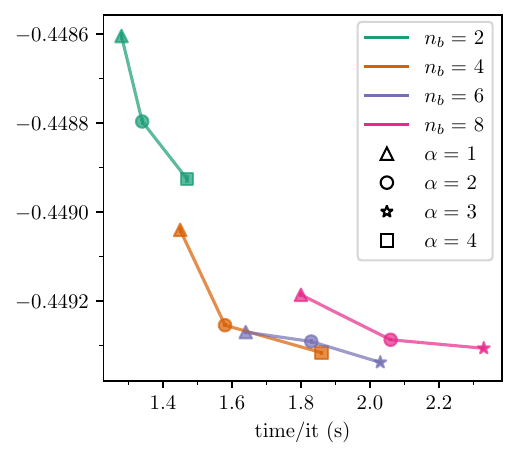}
    \end{minipage}
    \begin{minipage}{5cm}
    \includegraphics[width=\textwidth]{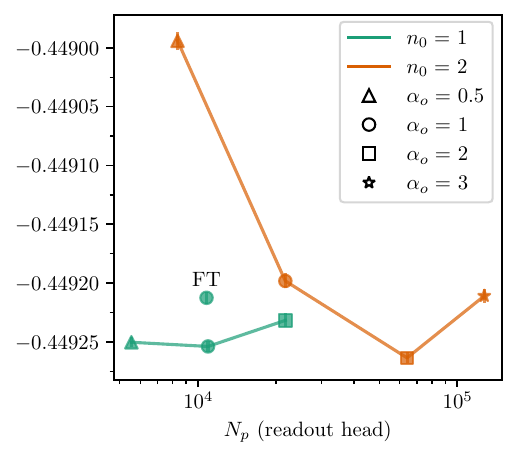}
    \end{minipage}
    \begin{textblock}{1}(0,-3)
		{\textbf{a.}}
	\end{textblock}
    \begin{textblock}{1}(5,-3)
		{\textbf{b.}}
	\end{textblock}
    \begin{textblock}{1}(9,-3)
		{\textbf{c.}}
	\end{textblock}
    \caption{Variational energies
    (\textbf{a.}) obtained for various $L$ with $(n_b,k,2)[1,1]$ networks, the dotted line is the energy obtained by the fViT in~\cite{viteritti2024}, the insets show the same data on a narrower energy scale for $n_b= 4,8$ and $12$.
    (\textbf{b.}) For $L = 10$, $(n_b, 3, \alpha)[1,1]$ networks, the time/iteration (x-axis) is measured for the first symmetry stage of the optimization.
    (\textbf{c.}) For $(4,3,2)[n_o,\alpha_o]$ networks (modifying the readout block), where $N_p$ is the number of parameters in the readout. 
    The number of parameters excluding the readout is $\sim 1\times10^5$.
    For the Fourier transform (FT) readout, the sum over patches is the last operation in the network, which allows one to project to $\mathbf{k} \neq 0$ sectors.}
    \label{fig:experiments}
\end{figure*}
%%%%%%

First, we compare the effect of patching versus not patching the input in the embedding layer.
When not patching, we still project to the relevant translational-invariant symmetry subsector in the readout (details are provided in appendix~\ref{app:unpatched}), therefore removing the inductive bias of the patching as well as treating inter and intra unit cell correlations on an equal footing, without modifying the symmetry of the wavefunction.

We perform a VMC calculation for an $L = 6$ system with a $(4,3,2)[1,1]$ network with and without patching, with results shown in fig.~\ref{fig:patching}.
We find that patching improves the accuracy of the final solution, reducing the relative error
\begin{equation}
    \epsilon = \abs{\frac{E - E_{\mrm{GS}}}{E_{\mrm{GS}}}},
\end{equation}
from $2.7\times 10^{-4}$ without patching, to $8.0 \times 10^{-5}$ with patching.
Furthermore, the patching reduces the cost of evaluating the network by approximately a factor of 4 (when using an equal number of features, see fig.~\ref{fig:summary}).

In a model, such as the $J_1$-$J_2$, where $2 \times 2$ patches are larger than the unit cell, enforcing the correct translational symmetry using Eq.~\ref{eq:trans_symm} requires multiple evaluations of the network, cancelling the reduced computational cost.
Nevertheless, lower energies are obtained with patching, likely due to the sublattice-dependent bias it provides, so proves to be a good design choice.
In the following we always use an embedding which patches the input.
%%%%%%%%%%%%%%%%%%%
\subsection{Convolutional kernel size and encoder depth}
We now turn to the encoder, first investigating how the convolutional kernel size $k$ and the encoder depth $n_b$ in $(n_b,k,2)[1,1]$ networks affect the accuracy of the calculation.
We benchmark on multiple system sizes $L = 6,8$ and $10$.
For each system size, we consider a kernel size up to $k=L/2$, at which point the convolutional kernel spans the entire patched lattice, so can encode global features in a single encoder block, as in the fViT.
Our results are summarized in fig.~\ref{fig:experiments}a.

\paragraph{Kernel size:}
For $L = 6$, the difference in energies obtained for $k=2$ and $3$ is comparable to numerical error from MC estimation of the energy ($\approx 5\times 10^{-6}$).
For $L = 8$ and $L = 10$, we observe that the largest kernel sizes ($k=4$ and $5$ respectively) perform worse than smaller kernel sizes as the networks become harder to optimize.
For all system sizes, the ConvNext achieves a comparable accuracy to the fViT.

Additionally, we find that the network depth should be chosen such that its receptive field allows for correlations among all lattice variables. 
For a small network depth ($n_b = 2$), using a kernel size $k=2$ limits the receptive field to $3 \times 3$, which does not span the entire lattice and therefore the output amplitude does not depend on long-range correlations of the input variables.
Increasing the kernel size to $k=3$ expands the receptive field to $5 \times 5$, allowing the encoder to capture all-to-all correlations within the lattice sizes we study. 
Indeed, for system sizes $L=8$ and $L=10$, this increase in kernel size significantly improves the variational energy, highlighting the importance of these global correlations. 
Furthermore, when we increase the network depth to $n_b=4$, all kernel sizes considered already have sufficiently large receptive fields to capture global correlations, which reduces the sensitivity of performance to the kernel size. 
Thus, a kernel size of $k=3$ consistently achieves optimal performance, and we adopt this choice in subsequent experiments.

\paragraph{Depth:}
We also observe a significant improvement in the variational energies when increasing the network depth from $n_b = 2$ to $4$, indicating that deeper networks are more effective. 
Further increases in depth beyond $n_b = 4$ continue to yield improvements, although these gains quickly diminish, becoming marginal when moving from $n_b = 4$ to $8$ and even smaller for greater depths.

\subsection{Encoder expansion ratio}
Next, we study the effect of modifying the expansion ratio $\alpha$ in the patch-wise non-linear feature transformation $g$ of the encoder (Eq.~\ref{eq:mlp}). 

Using $(n_b, 3, \alpha)[1,1]$ networks, we study the combined effect of modifying the encoder depth $n_b$ and $\alpha$, as summarized in fig.~\ref{fig:experiments}b.
We again observe a significant improvement in the energy when increasing $n_b$ from $2$ to $4$.
For $n_b \geq 4$ we find that it can be more efficient to increase the expansion ratio rather than the depth, achieving better energies at comparable computational cost.
For $n_b=6, \alpha = 4$ (not shown) our optimizations become unstable, whereas we deem the cost of an $n_b = 8, \alpha = 4$ network too high for our purposes.
Therefore, we take a $(6,3,3)$ encoder as the optimal choice.

As discussed further in Appendix~\ref{app:correlators}, $\alpha$ and $n_b$ play distinct roles in terms of how the hidden representation is constructed from correlators; $\alpha$ controls the number of different correlators computed each encoder block and $n_b$ the maximum number of recursions through the nonlinear block to construct correlators. 
Therefore one should tune both parameters in order to find a performant network design. 

We note that these values are close to the optimal choices $n_b=4, \, \alpha=4$ found for computer vision tasks~\cite{liu2022,woo2023}.

\subsection{Readout size}
Finally, we investigate the effect of modifying $n_o$ and $\alpha_o$ in the readout feed-forward network, with results summarized in fig.~\ref{fig:experiments}c.
By modifying the number of parameters in the readout, we are in principle modifying the representational power in mapping from the hidden representation constructed by the encoder to complex amplitudes.

We find that for a single layer readout, $\alpha_0$ makes little difference to the variational energy obtained, changing by only $\approx 2 \times 10^{-5}$ across all expansion ratios.
On the other hand, for two layers, the final energy changes significantly with the expansion ratio, although requires a relatively large number of parameters to perform on par with a single layer readout.
We also observe that the optimization becomes more challenging for $\alpha_o=3$, resulting in higher variational energies.
Nevertheless improvements in the energy when modifying the readout block are small suggesting that the power of the neural network to represent the ground state comes largely from the encoder block mapping to the hidden representation.

Despite this, the computational cost of increasing the number of parameters in the readout head is small, since it operates on input $\mrm X \in \mathbb{R}^{N_f}$ (see fig.~\ref{fig:summary}), so offers a computationally inexpensive way to improve variational energies.
We therefore choose a $[2,2]$ readout.

We also tested a readout layer based on the Fourier transform (FT) which could be used to project to $\mathbf{k} \neq 0$ sectors (see Appendix ~\ref{app:FT}).
The data corresponding to this approach, for the $\mathbf{k}=0$ mode, is included in Figs.~\ref{fig:experiments}c and~\ref{fig:summary}.
We observe that the energy is slightly higher than the standard readout layer based on a dense layer. 
This is likely due to the fact that the FT readout, the output is no longer explicitly a function of the sublattice magnetization (see Appendix \ref{app:sublattice}), which provides a useful bias for learning the sign structure of the ground state.
Nevertheless, one does not generally know in which momentum sector the ground state lies, so a thorough variational calculation should check several momentum sectors.

\subsection{System size dependence}
In fig.~\ref{fig:experiments}a, the difference between the ConvNext and fViT variational energies decreases as system size is increased.
This could be a result of different optimization procedures (for example, we use a fixed number of iterations across all system sizes), or a result of differences in the architectures.
Isolating the system-size scaling of the various architectures would require further numerical experiments where computational cost and optimization protocols are made equal. 
Note that when running for $L > 10$, we found the optimization to be rather unstable, which could likely be remedied by using transfer learning from smaller system sizes.

\subsection{Summary of design choices}
%%%%%%%%%%%%%%%
\begin{figure*}
\centering
\includegraphics[width = 18cm]{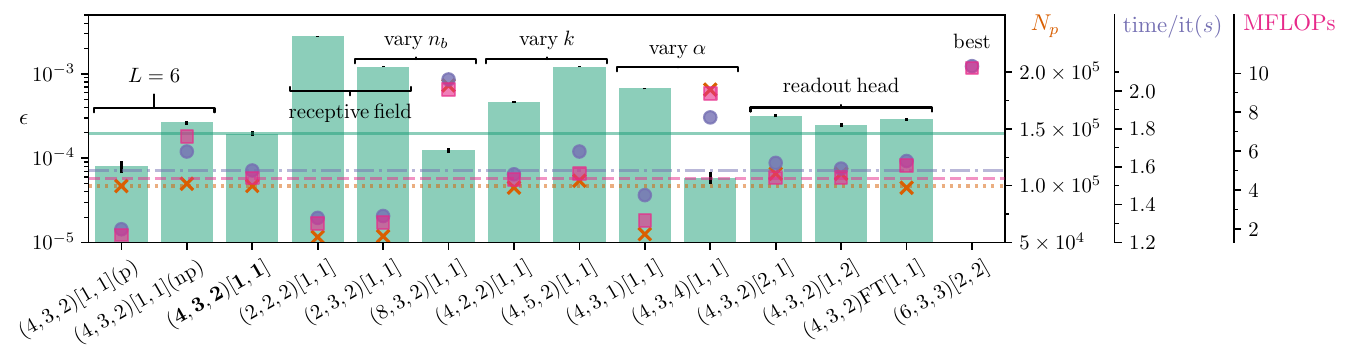}
\caption{Summary of the network changes and impact on performance. All results are for $L = 10$, except for the patching (p) and no patching (np) results which were obtained for $L = 6$. $\epsilon$ (bars) is measured relative to the ground state energy ($L = 6$) or our best variational energy ($L = 10$, see table~\ref{tab:benchmark}). The error bars are the MC error in the evaluation of the energy after optimization. Also plotted are $N_p$ (crosses), time/iteration of the optimization (circles) and MFLOPs for a single network evaluation (squares). The lines correspond to the values of these quantities for the reference $(4,3,2)[1,1]$ network.
}
\label{fig:summary}
\end{figure*}
%%%%%%%%%%%%%%%%
The results from the previous sections are summarized in fig.~\ref{fig:summary}, comparing the variational energies obtained against various measures of computational cost: number of parameters, $N_p$, total floating point operations (FLOPs) for a single evaluation of the network, as well as the time per iteration of the optimization.

Summarizing our findings on network design choices:
\begin{itemize}
    \item The receptive field should allow for all-to-all correlations, therefore the depth should be ${n_b = \mathrm{ceil}(\frac{L_i/P_i-k}{k//2}+ 1})$ where $L_i$ is the linear system size, $P_i$ the linear patch size.
    \item Larger kernel sizes are in general harder to optimize; $k=3$ works well and is a good compromise.
    \item There is an optimal combination of depth and expansion ratio which can be used to increase the size of the network. 
    \item A relatively small readout block is sufficient, but the number of parameters can be increased at little cost.
\end{itemize}

\subsection{Benchmark Results}
\begin{table}
	\small
	\centering
	\begin{tabular}{|c|c|c|c|}
	\hline
		Hamiltonian & Network & $E/N$ & $N_p$ \\
	\hline
	  Shastry-Sutherland & ConvNext & $-0.449342(4)$ & $2.6 \times 10^5$\\
	  ($J_1= 0.8,J_2 = 1$) & fViT\cite{viteritti2024} & $-0.449329(1)$ & $2.7 \times 10^5$  \\
      \hline
	  $J_1-J_2$ 		& ConvNext 	& $-0.497583(6)$ & $2.6 \times 10^5$\\
	  ($J_1 = 1, J_2 = 0.5$)	& fViT\cite{rende2023}    & $-0.497634(1)$& $2.7 \times 10^5$  \\
	  & ResNet\cite{chen2024a} &  $-0.4976921(4)$ & $1.07\times10^6$\\
	\hline
	\end{tabular}
	\caption{Benchmark variational energies for the Shastry-Sutherland and $J_1-J_2$ models, using  deep translationally-symmetry NQS.
    The corresponding number of parameters, $N_p$, are indicated.}
	\label{tab:benchmark}
\end{table}
Bringing all of the previous results together, we choose a network of $(6,3,3)[2,2]$ to compute energies for the $L = 10$ Shastry-Sutherland model and the $L = 10$ $J_1-J_2$ model.
Our results are summarized in table~\ref{tab:benchmark}, along with those from the ResNet in~\cite{chen2024a} and fViT~\cite{rende2023,viteritti2024}.
Variational energies differ with a relative error of at most $\approx 10^{-4}$.
These small differences can likely be accounted for by differing network sizes, number of samples and optimization procedures, rather than any inherent differences between architectures.
A more nuanced comparison would also compare total computational cost of the calculations, across several different Hamiltonians and lattice sizes. 
\section{Discussion}
\label{sec:discussion}
As we have shown when introducing the ConvNext and in the results we obtain on benchmark problems, successful translationally-symmetric NQS employed so far in the literature operate under the same principles, with different architectures offering more or less efficiency or inductive bias for the problem at hand.
For the two benchmark problems we study here, where translationally-symmetric NQS obtain state of the art variational energies, the ground states are thought to be in the same phase (a $\mathbb{Z}_2$ Dirac spin liquid~\cite{maity2024}), as well as having similar Hamiltonians, so it is unclear if these architectures are specifically well-suited to tackle these problems, or can successfully be applied to other ground states and Hamiltonians.
As we have mentioned, the inductive bias of the embedding is well-suited to learn the sign structure of states which (approximately) follow the Marshall sign rule, but for other models and lattice geometries the ground state sign structure may be far from such a sublattice-dependent rule.

We have focused on translationally-symmetric architectures as they can naturally include properties of condensed matter systems, translational symmetry and locality, as well as allowing one to project to different momentum sectors without incurring a cost which scales as the size of the system.
As seen in fig.~\ref{fig:patching}, including point group symmetries is also crucial for obtaining good variational energies with these networks, which is performed ``externally" by quantum number projection~\cite{mizusaki2004,nomura2021b} (see appendix~\ref{app:vmc}).
Group convolutional neural networks offer a more natural way to handle both translational and point group symmetries within the network, yet on, for example the $J_1$-$J_2$ model, are outperformed by translationally-symmetric architectures.
This may be due to the fact that imposing point group symmetries via quantum number projection allows one to introduce point group symmetries over the course of the optimization rather than from the outset, reducing the computational cost and allowing one to optimize large networks.
More speculatively, it may be that by not enforcing these symmetries from the outset, the larger Hilbert space available earlier in the training helps the network to ultimately find a better solution~\cite{wilson2025}.

Lattice symmetry-agnostic architectures such as recurrent neural networks~\cite{hibat-allah2020,hibat-allah2021a,hibat-allah2023b,moss2025} and transformers~\cite{sprague2024a,chen2025} have advantages in principle such as direct sampling, or a self-attention mechanism to learn relevant correlators, yet these do not appear to outperform translationally-symmetric architectures on frustrated benchmark problems in practice.

More biased approaches may be used when sufficient information about the problem is known, for example exploiting the emergent gauge symmetries of quantum spin liquids~\cite{kufel2025}, yet may not be directly applicable when the nature of these emergent symmetries is not already known. 
One can also combine fermionic mean-field ans\"atze with NQS~\cite{nomura2017,nomura2021}, which has yielded accurate variational energies, though computations are expensive.
This cost could be partially reduced, and expressivity of the ansatz increased, by combining with translationally-symmetric neural networks. 

Overall, translationally-symmetric architectures appear to provide useful inductive biases to study frustated ground states, whilst allowing for the use of large networks and lattice symmetries to improve variational energies.
\section{Conclusion and Outlook}
We have applied the ConvNext architecture as an NQS, using VMC for the Shastry-Sutherland model to optimize design choices, providing a blueprint for choosing network hyperparameters and clarifying the role of the various components of translationally-symmetric architectures.
Using our optimal network on both $J_1$-$J_2$ and Shastry-Sutherland models we find variational energies which are comparable to the lowest in the literature.
We discuss how the ConvNext is related to other deep translationally equivariant NQS, providing a unified perspective on the fViT and convolutional networks, such as ResNets.

In future work, the ConvNext architecture could be readily applied to models of 2D frustrated magnets, with and without multi-site unit cells, potentially helping to resolve cases where ground state phase diagrams are debated.
Furthermore, the architecture can be relatively straightforwardly generalized to three dimensional systems, by replacing two-dimensional convolutions with three-dimensional ones.
Although translationally-equivariant architectures have yielded remarkable results on various benchmark problems, there is not yet a systematic understanding of what kinds of states and Hamiltonians they are suited to, which is important both for understanding their domain of applicability and for understanding how to modify architectures to tackle a broader range of problems.

Beyond ground states, using projection to non-zero momenta, translationally-equivariant architectures can be used to perform a limited form of spectroscopy, whereby one can obtain an approximation of the lowest energy state in each symmetry sector of the Hamiltonian using VMC~\cite{morita2015a,nomura2021b,nomura2021,roth2023a,chen2024a}.
There has also been progress in using NQS to simulate dynamics~\cite{carleo2017a,schmitt2020,gravina2024,sinibaldi2024,walle2024}  and obtaining spectroscopic quantities such as the dynamical structure factor~\cite{mendes-santos2023}, which can be probed in solid-state experiments using, for example, neutron scattering.
Computing such quantities using NQS would be useful for reconciling experimental observations with predictions from theoretical models, providing a path to unambiguous experimental identification of exotic magnetic states of matter, such as quantum spin liquids.

\section*{Data and Code Availability}
Code for the implementation and optimization of the ConvNext is available at~\cite{code} and all data supporting the findings of this article at~\cite{data}.

\section*{Software}
Simulations were performed with NetKet \cite{netket3:2022,netket2:2019}, and at times parallelized with mpi4JAX \cite{mpi4jax:2021}.
This software is built on top of JAX \cite{jax2018github} and Flax \cite{flax2020github}.

\begin{acknowledgments}
We acknowledge extensive discussions with L.L. Viteritti and R. Rende.
The authors acknowledge support by the French Agence Nationale de la Recherche through the NDQM project, grant ANR-23-CE30-0018.
This project was provided with computing HPC and storage resources by GENCI at IDRIS thanks to the grants 2023-AD010514908 and 2024-A0170515698 on the supercomputer Jean Zay's V100/A100 partition.
\end{acknowledgments}

\appendix
\section{Variational Monte Carlo with Neural Network Quantum States}
\label{app:vmc}
In order to approximate ground states of quantum many-body spin Hamiltonians, we use VMC~\cite{becca2017}, optimizing our ansatz with the SPRING algorithm~\cite{goldshlager2024}, based on minSR~\cite{chen2024a, rende2023}.

Consider the quantum state, $\ket{\Psi_{\bm{\theta}}}$ with $N_p$ parameters $\bm{\theta} = (\theta_0,\dots,\theta_{N_{p-1}}) \in \mathbb{R}^{N_p}$ .
Expanding in the local $\hat{\sigma}^z$ basis, $\bm{\sigma} = (\sigma_0^z,\dots,\sigma_{N-1}^z) \in \mathbb{Z}_2^N$, for $N$ degrees of freedom,
\begin{equation}
    \ket{\Psi_{\bm{\theta}}} = \sum_{\bm{\sigma}} \braket{\bm{\sigma}}{\Psi_{\bm{\theta}}} \ket{\bm{\sigma}} = \sum_{\bm{\sigma}} \Psi_{\bm{\theta}}(\bm{\sigma}) \ket{\bm{\sigma}}.
\end{equation}
The energy of the variational state
\begin{equation}
    E[\Psit] = \frac{\bra{\Psit} \hat{H} \ket{\Psit}}{\braket{\Psit}{\Psit}} = \sum_{\bsigma} \frac{\abs{\Psit(\bsigma)}^2}{\braket{\Psit}{\Psit}} E_{\mrm{loc}}(\bsigma)
\end{equation}
is estimated via Monte Carlo sampling from $p(\bsigma) = \abs{\Psit(\bsigma)}^2$,
\begin{equation}
    \bar{E}[\Psit] = \frac{1}{N_s} \sum_{i=1}^{N_s} E_{\mrm{loc}}(\bsigma_i),
\end{equation}
where
\begin{equation}
    \bm v_i = E_{\mrm{loc}}(\bsigma_i) = \sum_{\bsigma'} \bra{\bsigma_i} \hat{H} \ket{\bsigma'} \frac{\Psit(\bsigma')}{\Psit(\bsigma_i)}.
\end{equation}

In VMC we want to find the $\btheta$ which minimizes the energy, thus finding an approximation of the ground state. 
We use the SPRING algorithm (without norm constraint) to optimize the parameters, which are updated at each step as $\btheta_{k+1} = \btheta_k + \Delta \btheta_k$, with
\begin{equation}
    \Delta \btheta_k = \eta\bigg[ \mrm{X}^T \bigg(\mrm X \mrm X^T + \lambda \mathbb{I} \bigg)^{-1} \tilde{\bm{v}} +p\Delta\btheta_{k-1}\bigg],
\end{equation}
where 
\begin{align}
    \tilde{\bm{v}}_i &= \mrm{Re}(\bm{v}_i)- p(\mrm{X}\Delta\btheta_{k-1})_i,\\
    \tilde{\bm{v}}_{i+N_s} &= \mrm{Im}(\bm{v}_i)- p(\mrm{X}\Delta\btheta_{k-1})_{i+N_s},\\
    \mrm{X}_{i,j} &= \frac{\mrm{Re}(\mrm O_{i,j} - \bar{\mrm O}_i)}{\sqrt{N_S}},\\
    \mrm{X}_{i+N_s,j} &= \frac{\mrm{Im}(\mrm O_{i,j} - \bar{\mrm O}_i)}{\sqrt{N_s}},
\end{align}
with $\tilde{\bm{v}} \in \mathbb{R}^{2N_s}$ and $\mrm{X} \in \mathbb{R}^{2N_s \times N_p}$ which is computed from the Jacobian
\begin{equation}
    \mrm O_{i,j} = \frac{\partial \log \Psit(\bsigma_i)}{\partial \theta_j}.
\end{equation}
Hyperparameters for the optimization are the learning rate, $\eta$, diagonal shift, $\lambda$ and momentum $p$.
For $p = 0$, the algorithm reduces to minSR.

In addition to the translational invariance which we impose within the neural network, we also apply quantum-number projection~\cite{mizusaki2004,nomura2021b,reh2023} for the point group (PG) and spin-parity (P) symmetries, resulting in the symmetrized ansatz
\begin{equation}
    \Psit(\bsigma)_{\mrm{symm}} = \frac{1}{\abs{G}} \sum_{g \in G} \chi_g \Psit(g^{-1} \bm{\sigma}),
    \label{eq:symmetrize}
\end{equation}
where $G$ is the group made up of elements $g$, which is being symmetrized over, and, for simplicity, we restrict ourselves to the invariant irreducible representation, with character $\chi_g = 1, \forall g \in G$.
Further details of the optimization procedure are provided in appendix \ref{app:optimization}.

\section{Components of translationally-symmetric neural quantum states}
\label{app:trans_symm}

\subsection{Sublattice bias from patching}
\label{app:sublattice}
Here we discuss how the patched embedding constructs a hidden representation in terms of sublattice-dependent quantities, which, for example, could allow the network to learn the Marshall sign rule~\cite{marshall1955,szabo2020} on a bipartite lattice.

If we were to pass the output of the embedding directly to the readout block (see Eq.~\ref{eq:stem_hidden}), followed by a sum pooling over the patches, the hidden representation is given by
\begin{equation}
	\tilde{\sigma}^{\mu} = \sum_x \mrm W^{\mu}_x m^z_x,
\end{equation}
with the sublattice magnetization
\begin{equation}
	m^z_x = \sum_r \sigma^z_{r,x}
\end{equation}
and the coefficients of $\mrm W$ determine how much the magnetization contributes to that specific feature.
Then passing through the remainder of the readout block, the full wavefunction encoded is
\begin{equation}
	\Psi_{\theta}(\bm{\sigma}) = g^o\bigg(\sum_{x} W_x^{\mu} m^z_x \bigg).
\end{equation}
The Marshall sign rule specifies that the sign of the amplitude $\Psi(\bm{\sigma})$ is given by $(-1)^{n_{\uparrow,x}}$,
where $n_{\uparrow,x}$ is the number of $\sigma^z = +1$ spins on the $x$ sublattice, such that
\begin{equation}
	n_{\uparrow,x} = \frac{m^z_x + N_x}{2}
\end{equation}
where $N_x$ is the total number of sites on the $x^{\mrm{th}}$ sublattice.
Thus, assuming the readout block is sufficiently expressive, the network can learn the sign rule, as a result of the sublattice-dependent bias introduced by the patching.
In a network with a skip connection the output of the embedding block is indeed passed directly to the readout with further corrections from the encoder, which explains why architectures which patch the input do not require the Marshall sign rule as a prior~\cite{viteritti2024}.

\subsection{Generating higher-order correlators from convolutional layers}
\label{app:correlators}
To understand how encoder blocks in translationally-equivariant architectures construct hidden representations of the $\sigma_i^z$ variables, let us consider how standard convolutional layers transform the $\sigma_i^z$ variables.
This will also allow us to understand the roles of the various components of the ConvNext encoder.

Consider a single-feature, feed-forward layer, which has a receptive field of $n$ variables, $\tilde{\sigma}_i^z$.
For simplicity, we also consider a system with a single spatial dimension, such that the effect of the layer is
\begin{equation}
	y = f\bigg(\sum_{i=0}^{n-1} w_i \tilde{\sigma}_i^z + b \bigg) = f(\tilde{x}+b),
\end{equation}
where $f$ is a non-linear activation function.
Assuming $f$ is analytic and has a convergent Taylor series, then expanding around $b$ to order $k_{\mrm{max}}$~\cite{denis2024},
\begin{equation}
	y = \sum_{k=0}^{k_{\mrm{max}}} \tilde{x}^k\frac{f^{(k)}(b)}{k!} + R_{k_{\mrm{max}}}(\tilde{x}+b),
\end{equation}
where $R_{k_{\mrm{max}}}$ is a remainder term and the $\tilde{x}^k$ terms are $k^{\mrm{th}}$ order correlators
\begin{equation}
	\tilde{x}^k \propto \tilde{\sigma}^z_{i_0}\tilde{\sigma}^z_{i_1} \dots \tilde{\sigma}^z_{i_{k-1}} + \dots .
\end{equation}

Since a standard convolutional layer (Eq.~\ref{eq:convolution}) mixes both feature and position degrees of freedom, the kernel weights select the relative contributions of the variables within the correlators, with access to all feature channels.
The convolutional kernel width defines the receptive field, setting the maximum distance between variables which can be included.
The recursion of convolutional layers allows these correlators to be further combined, whilst skip connections retain the correlators generated by previous layers.

In the ConvNext  (eq.~\ref{eq:convnext-layer}) the standard convolution is replaced by a depth-wise convolution, $\mrm K$, which restricts which input variables each feature channel has access to.
The expansion matrix, $\mrm W$, chooses linear combinations of variables within each patch from which the correlators should be constructed, with the expansion ratio, $\alpha$, controlling the number of different correlators to be generated by the non-linearity.
Subsequently, the compression matrix, $\mrm {V}$, chooses linear combinations of the generated correlators to be carried forward. 

\subsection{Factored multi-head attention vs depth-wise convolutions}
\label{app:fmha_comp}
\begin{figure}
\centering
\includegraphics[width = 8cm]{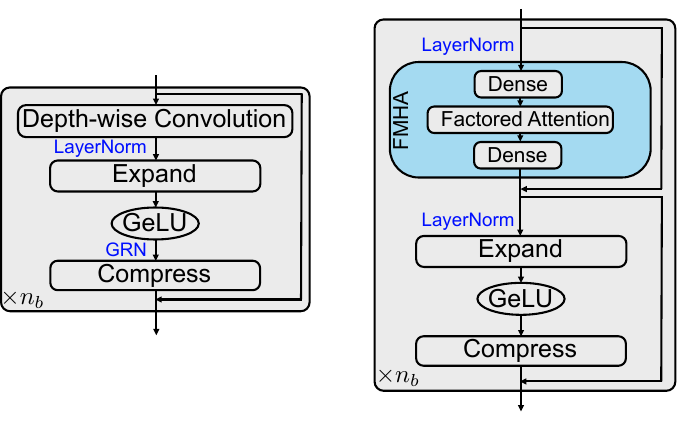}
\begin{textblock}{1}(-0.5,-2.5)
    {\textbf{a.}}
\end{textblock}
\begin{textblock}{1}(2.9,-2.5)
    {\textbf{b.}}
\end{textblock}
\caption{Comparison of encoder blocks in the ConvNext (\textbf{a.}) and fViT(\textbf{b.})\cite{ViT_implementation}.
In the ConvNext, correlations between different patches are created by a depth-wise convolution, whereas in the fViT this is done by factored multi-head attention (FMHA).
When the number of heads in the fViT is equivalent to the number of features the factored attention operation is equivalent to a depth-wise convolution.
}
\label{fig:convnext_vit_comparison}
\end{figure}

The key conceptual difference in the  fViT compared to convolutional networks is that in its embedding block, the data is first reshaped (patched) to $\mathbb{R}^{P \times N_x}$, where $P = P_xP_y$, and $N_x$ is the total number of sites in a patch.
That is, the patches are flattened along a single-dimension, and the data passes through the network with this shape.
In this form, it is not so easy to see how the operations relate to convolutional networks, where one preserves the position dimensions.
In the following we translate the factored multi-head attention (FMHA) block (see Fig.~\ref{fig:convnext_vit_comparison}) to operation on $\tilde{\sigma} \in \mathbb{R}^{P_x \times P_y \times N_f}$ input, in order to make the connection to convolutions transparent.
%%%%%%%%%%%%%%%
%%%%%%%%%%%%%%%%
Consider a factored ViT where the number of heads, $N_h = N_f$, then the factored multi-head block,  is as follows.
First, a dense layer is applied over the features,
\begin{equation}
	\tilde{\sigma}_{i,j}^{'\nu} = \sum_{\mu=0}^{N_f-1}  \mrm V^{\nu,\mu}  \tilde{\sigma}_{i,j}^{\mu},
	\label{eq:head_dense}
\end{equation}
where $\mrm V$ is an $N_f \times N_f$ matrix of parameters.
Then, in order to preserve translational equivariance the factored attention mechanism~\cite{rende2024b} is used,
\begin{equation}
	\tilde{\sigma}_{i_2,j_2}^{'\mu} = \sum_{i_1,j_1} \mrm J_{(i_2,j_2),(i_1,j_1)}^{\mu} \tilde{\sigma}_{i_1,j_1}^{\mu}
	\label{eq:factored_attention}
\end{equation}
where 
\begin{gather}
	\mrm J_{(i_2,j_2),(i_1,j_1)}^{\mu} = \mrm J_{(i'_2,j'_2),(i'_1,j'_1)}^{\mu}  \nonumber \\ \mathrm{if} \: \mathbf{x}_{i_2} - \mathbf{x}_{i_1} + \mathbf{y}_{j_2}  - \mathbf{y}_{j_1} = \mathbf{x}_{i'_2} - \mathbf{x}_{i'_1} + \mathbf{y}_{j'_2}  - \mathbf{y}_{j'_1}, 
\end{gather}
that is $\mrm J$ is translationally invariant, it has the same value when mapping between input and output patches with the same relative displacement.
This is followed by another dense layer in the form of eq.~\ref{eq:head_dense}.

Due to this translationally invariant structure, this is equivalent to the depth-wise convolution,
\begin{equation}
	\tilde{\sigma}^{'\mu}_{i,j} = \sum_{m,n} \mrm K_{m,n}^{\mu} \tilde{\sigma}^{\mu}_{i+m,j+n},
\end{equation}
where the kernel size is $P_x \times P_y$, since a convolutional kernel also has the property that weights connecting input and output positions are the same if the relative displacement between the two is the same.

Decreasing the number of heads $N_h < N_f$ in factored multi-head attention (as is done in practise) introduces parameter sharing over the features, with $N_h$ distinct $\mrm J_{(i_2,j_2),(i_1,j_1)}$ in eq.~\ref{eq:factored_attention}.

Since the ConvNext encoder uses a depth-wise convolution followed by expand and compress operations, whereas the fViT encoder uses factored multi-head attention also followed by expand and compress operations, we see that these architectures are similar.
In our experiments, we find that kernel sizes which span the system, e.g for $k = 5$ in fig. \ref{fig:experiments}a, can be difficult to optimize, which is not the case for the fViT.
This may be a result of parameter-sharing across features in the attention or the dense layers before and after, see Fig.~\ref{fig:convnext_vit_comparison}, helping to stabilize larger kernels (we verified that introducing two skip connections per encoder block in the ConvNext was not able to stabilize these large kernels). 

\subsection{Projection to Momentum Sectors}
\label{app:FT}
In order to project the wavefunction to $\mathbf{k} \neq 0$ sectors, the sum over patches (see Fig.~\ref{fig:convnext}) is moved to the end of the readout head and generically replaced with (see Eq. ~\ref{eq:trans_symm} and ~\cite{chen2024a})
\begin{equation}
    \log \Psi_{\theta}^{\mathbf{k}}(\bm{\sigma}) = \log \sum_{i=0}^{N_p-1} e^{-i\mathbf{k} \cdot \mathbf{r}_i} \Phi_{\theta,i}(\bm{\sigma}),
\end{equation}
where $\Phi_{\theta,i}(\bm{\sigma})$ is the output of the network for the patch $i$ at position $\mathbf{r}_i$.
We find that for the $\mathbf{k} = 0$ sector, this slightly increases the energy obtained, likely due to the fact that the output of the network is not explicitly a function of $m^z_x$ (see appendix~\ref{app:sublattice}).
However, it is generally desirable to obtain results in different $\mathbf{k}$ sectors, as the ground state need not generically be in the $\mathbf{k} = 0$ sector and also gives information on the low-energy spectrum of the model.

\section{Optimization Procedure}
\label{app:optimization}
Our optimization protocol makes use of different symmetry stages~\cite{rende2023} as shown in fig.\ref{fig:patching}, projecting to the invariant representation of all symmetries.
We begin the optimization with the bare network, which internally is made to be translationally invariant.
We then introduce point group and spin-parity externally using eq.~\ref{eq:symmetrize}.
First, we symmetrize over the rotational elements of the point group, that is $C_4$ rotations.
We then symmetrize over the full point group, adding reflection in the line $x = y$ to the group to be symmetrized over.
Finally we add spin parity symmetry, under $\hat{P} = \prod_i \hat{S}_i^x$, to the symmetrization.
The advantage of this approach over symmetrizing at the outset is two-fold; optimization is computationally cheaper the fewer symmetries we enforce externally, as well as being more stable.

Each iteration, we use the SPRING algorithm described in section~\ref{app:vmc} to update the parameters.
We found that using $p = 0.9$ led to faster convergence when comparing different $p$ over the first $1000$ iterations, so use this value.
In the numerical experiments for designing the network we typically perform $2500$ iterations in each symmetry stage, with a cosine decaying learning rate~\cite{loshchilov2017} from $10^{-2}$ to $5\times10^{-3}$ and a cosine decaying diagonal shift from $10^{-2}$ to $10^{-6}$ in each stage.
We use a total of $4092$ samples per iteration, with $512$ MC chains on each of $8$ A100 GPUs ($1$ sample per chain), each chain performing a single MC sweep of $N$ MC steps.

At the end of the optimization, we freeze the learning rate and diagonal shift, performing $50$ additional optimization steps with more samples per chain, discarding half for thermalization.
We use $32$ chains per GPU and $8196$ total samples.
To compute an expectation value for the energy we then load in the parameters which gives the minimum energy over these $50$ iterations and use $6.4\times 10^5$ samples on $32$ chains on a single GPU (half of the samples are discarded for thermalization).
\section{Unpatched Architecture}
\label{app:unpatched}
In order to assess the impact of patching on network performance, we modify the ConvNext architecture to remove patching whilst still maintaining translational invariance up to the unit cell (of the Shastry-Sutherland model). 
To this end, we remove the embedding block and increase the kernel size of the first convolution of the encoder, in order to have the same receptive field in both cases.
Therefore the data passing through the encoder is of form $\mathbb{R}^{L_x\times L_y \times N_f}$, with spatial dimensions corresponding to the sites of the lattice.

In the readout head we perform a sum over the $b^2$ basis sites in the unit cells
\begin{equation}
    \tilde{\sigma}^{'\mu}_{i,j} = \sum_{n,m=0}^{L/b-1} \tilde{\sigma}_{i+nb,j+mb}^{\mu}
\end{equation}
to take the encoder output to the form $\mathbb{R}^{b \times b \times N_f}$.
We then flatten the spatial dimensions and perform a dense layer
\begin{equation}
    \tilde{\sigma}^{'\mu} = \sum_{i=0}^{b^2-1} \mrm W_i \tilde{\sigma}_{i}^{\mu},
\end{equation}
equivalent to a strided convolution of stride $b$, taking to the form $\mathbb{R}^{N_f}$, after which the data passes through the remainder of the usual readout block.
If the problem at hand is a lattice with a single-site unit cell, the appropriate translational symmetry can be enforced by replacing the above two operations with pooling over sites or using a readout block which can project to $\mathbf{k} \neq 0$ sectors (Appendix~\ref{app:FT}).

\bibliography{bibliography}

\end{document}